\documentstyle[11pt,newpasp,twoside,psfig]{article}
\def\edcomment#1{\iffalse\marginpar{\raggedright\sl#1\/}\else\relax\fi}
\reversemarginpar
\begin{document}
\title{Density waves in the central regions of galaxies}
 \author{Eric Emsellem}
\affil{Centre de Recherche Astronomique de Lyon, 9 av. Charles Andr\'e,
	69561 Saint-Genis Laval Cedex, France}

\begin{abstract}
Density waves in the central kpc of galaxies,
taking the form of spirals, bars and/or lopsided density
distributions are potential actors of the redistribution of
angular momentum. They thus play an important role
in the overall evolution of the central structures, not mentioning
the possible link with the active/non-active nucleus.
I present here kinematical evidences for the presence of such structures
using new sets of observations: two-dimensional (OASIS/CFHT)
and long-slit (ISAAC/VLT) spectrography of nuclear bars and spirals.
I also discuss the importance of $m=1$ modes in the nuclear regions
of galaxies, illustrating this with newly revealed cases and
original N body simulations.
\end{abstract}

\section{Introduction}
Density waves are now recognised as important actors in the evolution
of the internal structures of galaxies. This is often emphasized via their
role in the redistribution of angular momentum. Since the contributions
of Lindblad, Lin and Shu, astronomers have gradually associated observed 
spirals, bars and (more seldom) warps to (kinematic) density waves. 
Lopsidedness has however been mostly overlooked as a possible mode for
density waves (but see e.g. Rudnik \& Rix 1998,
Swaters et al. 1999, Tremaine 2001). In this paper,
we will present kinematical evidences for so-called $m=2$ (spirals and bars)
and azimuthal $m=1$ modes in the central part of galaxies, thus
occuring at scales from a few parcsecs to a few hundreds of parsecs.

\section{Nuclear spirals ...}
A number of nuclear spiral-like dust and/or gas structures were recently observed
in disc galaxies mostly using the high resolution imaging capabilities
of HST (e.g. Regan \& Mulchaey 1999; Martini \& Pogge 1999; Laine et al. 1999;
Chapman, Morris \& Walker 2000). These are however usually low-contrast
features for which, until now, only weak streaming motions were registred.
Emphasizing the fact that gas spiral density waves can be present
at all radii (contrarily to stellar density waves), Englmaier \& Shlosman (2000)
presented two-dimensional numerical simulations which, they argue, could account
for these low arm-interarm contrast spirals in the central parts of galaxies. These
would then represent the central extension of large-scale spirals
through the Inner Lindblad Resonance.
\begin{figure}
\centerline{\psfig{figure=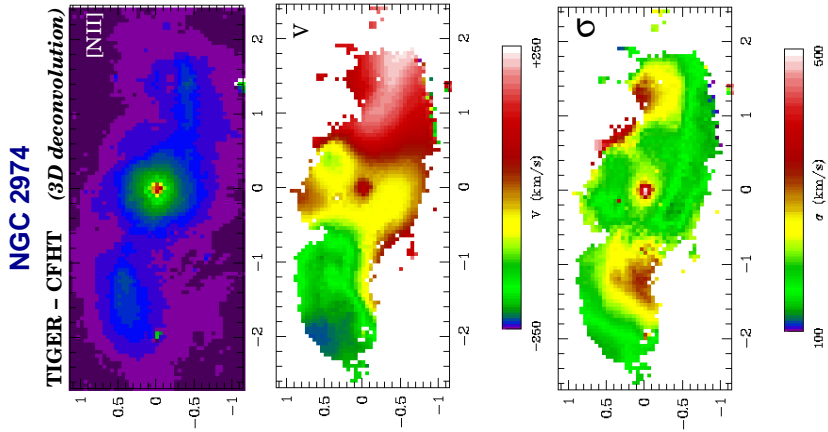,width=11cm,angle=90}}
\caption[]{Emission line distribution, velocity
and velocity dispersion maps obtained from a deconvolved TIGER/CFHT datacube.}
\end{figure}

In the course of a limited observational program aimed at studying
the coupling between the stellar and gaseous components in early-type
galaxies, Paul Goudfrooij and I have discovered a two-arm gas spiral within
the central 250~pc of an elliptical, NGC 2974, using two-dimensional
TIGER/CFHT spectroscopic data and subsequent WFPC2/HST images. The 
integral field datacubes allowed us to build the full stellar and gas velocity and dispersion
maps (Emsellem \& Goudfrooij, 2001, in preparation):
the observed gas kinematics reveals very strong streaming motions and hints
for shocks on the back side of the trailing spirals (Fig.~1).
We have modeled the gravitational potential using the photometry and stellar kinematics
as constraints, and then attempted to fit the observed gas velocities
by using the formalism developed by Shu et al. 1973. Our best fit model
requires the ILR to be within $\sim 70$~pc of the centre, which coincidently 
corresponds to the last radius until which we can follow the spiral.
This nuclear spiral is not self-gravitating, but it is 
not of the same nature than the ones described in Englmaier \& Shlosman (2000).
Refined models and accurate infrared photometry should now be used
to more accurately determine the characteristics of this density wave
and the triggering source (tumbling potential).

\section{Double bars ...}

Didier Greusard, Daniel Friedli and I have also embarked in a program to retrieve the kinematics
of nuclear bars in double barred systems, with the aim of asserting
the physical reality of such structures, and of understanding their
role in the redistribution of the dissipative component.
We have observed a number of double bars using the integral field spectrograph
OASIS at the CFHT. We derived the gas and stellar
kinematics of a few nuclear bars, to be compared with original section N body $+$ SPH
simulations by Daniel Friedli. In the case of NGC~2859,
we could fit the kinematical maps only by including a secondary decoupled nuclear bar in the model. 
We can now use these models to constrain the pattern speed of the two bars
as well as the gas inflow rate.
\begin{figure}
\psfig{figure=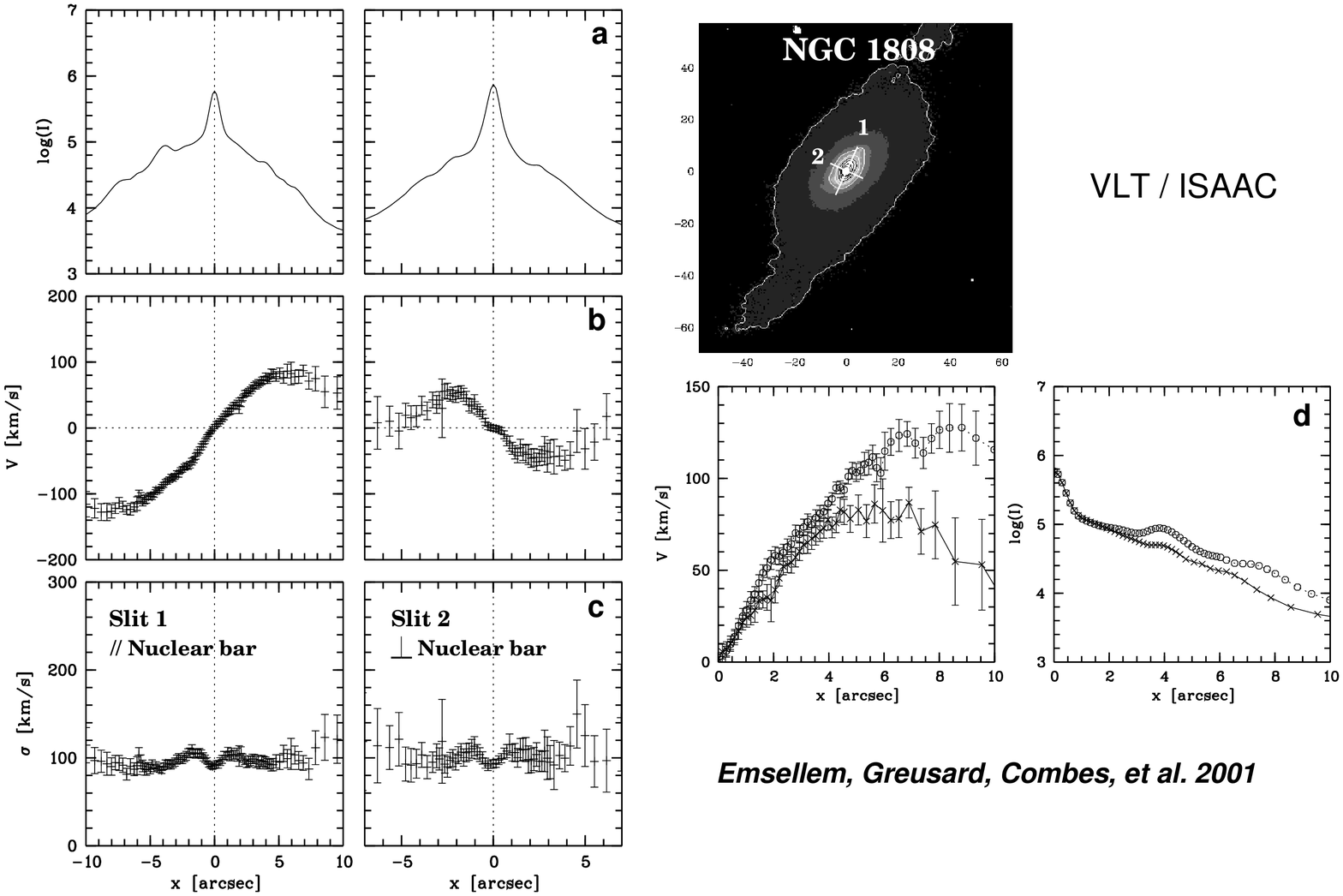,angle=90,width=14cm}
\caption[]{The top right panel shows the
NIR image of the central part of N1808, 
with the positions and lengths of the two slits, parallel and perpendicular to the nuclear bar
(the labels 1 and 2 overimposed on the image indicate positive abscissa);
From panel a to c: the luminosity profiles (in log) along the two slits,
velocity and dispersion profiles. Panels d shows the folded luminosity and velocity
profiles along the major-axis.}
\label{fig:n1808}
\end{figure}

We have also conducted, via a larger collaboration (including Fran\c{c}oise Combes, 
St\'ephane Leon, Emmanuel
P\'econtal and Herv\'e Wozniak), a similar study this time using near-infrared long-slit
spectroscopy to trace the stellar kinematics in later type barred spirals,
and examine the link between the bar and the nuclear activity (Emsellem et al. 2001).
We have obtained the kinematical profiles for 3 active and 1 starburt galaxies
along the major and minor axes of the secondary bar.
These data reveals a clear kinematical decoupling at the outer edge
of the inner bar, thus confirming that the nuclear bars revealed in the photometry are truly
decoupled dynamical entities. But the most surprising result comes from
the stellar velocity dispersion (Fig.~2): 3 out of the 4 galaxies exhibit a significant
drop in the central $\sigma$ profiles (the Seyfert~1 nucleus of the fourth galaxy
prevented us to derive any meaningful stellar kinematics within the central half arcsecond).
We interpret this as evidence for the presence of a cold transient component, recently formed via
accretion of gas. This scenario is supported by our study of the luminosity
weighted mean stellar population (Greusard et al., in preparation), and new
N body + SPH simulations including star formation (Wozniak et al., in preparation). 
In order to further comment on the link between this late accretion
and the nuclear activity new ISAAC/VLT data on inactive and/or
unbarred galaxies are required.

\section{... and $m=1$ modes}
\begin{figure}
\psfig{figure=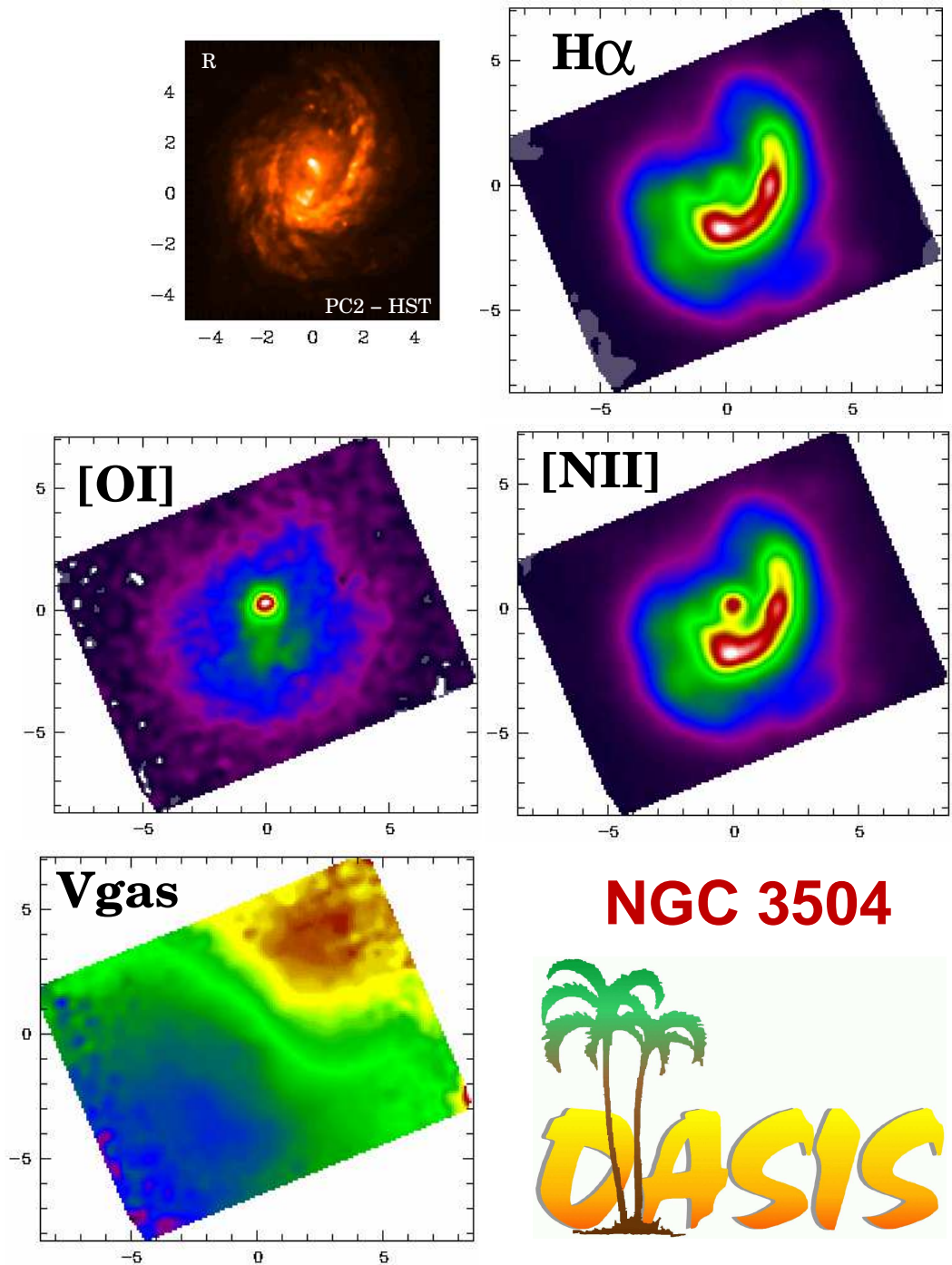,width=14cm}
\caption{OASIS/CFHT emission line distribution and velocity maps
of NGC~3504, a double barred galaxy. The top left panel is a comparison
WFPC2 image.}
\label{fig:n3504}
\end{figure}

Lopsided luminosity/mass distribution are often apparently superimposed on $m=2$ modes
(see e.g. Kamphuis et al. 1991). But as for $m=2$ modes, these should
be confirmed via some kinematical signatures. An example of such a detection is illustrated
in panel d of Fig.~2 where the radial surface brightness and velocity profiles
of NGC~1808 are shown: there is a clear difference between the two sides both in 
the photometry and in the kinematics. Another beautiful illustration of such asymmetries
is present in the central part of the double barred galaxy NGC~3504:
the apparent asymmetric luminosity distribution corresponds to a real offset
as emphasized in the adaptive optics K band image obtained by Combes and collaborators.
In the visible, this appears as a one arm spiral, clearly delineated in the gas distribution
derived from OASIS/CFHT datacubes (Fig.~3). The higher density gas
traced by the [OI] emission line exhibits a different distribution, with
large streaming motions between the centre and the edge of the spiral.
This could be the signature of infalling gas, although this should be confirmed
by proper hydrodynamical simulations.

A final example at a much smaller scale (of a few pc) is provided by the nucleus of M~31.
We direct the reader to our recent paper (Bacon et al. 2001) for details on this
object. The main issue is the observed asymmetry in the photometry, which exhibits
an extra luminosity peak called P1, offset from the apparent centre P2 of the outer isophotes.
Tremaine (1995) endorsed the presumed presence of a supermassive black hole near P2, 
then suggesting that the dynamical cold P1 could be explained if eccentric keplerian orbits
are gradually aligned via the dynamical friction onto the slowly rotating bulge stars.
Dynamical friction was found to be negligible (Emsellem \& Combes 1997,
Bacon et al. 2001), but we managed to beautifully fit the lopsided surface brightness distribution
by self-consistent N body simulations of a nuclear stellar disk suffering an $m=1$ mode.
This is a very slow mode with $\Omega_p \sim 3$~km.s$^{-1}$.pc$^{-1}$, a factor of about
one hundred slower than the local circular velocity, and it can remain for hundreds
of dynamical times. Such a mode seems to grow spontaneously, given that the disk
is about 20-40\% of the mass of the central black hole, or could be triggered
by some perturbers (infalling gas cloud). This may thus be a recurrent process,
and could well be a common phenomenon among galaxy nuclei.

\section{Conclusion}
In this paper, we have shown a few illustrations of $m=2$ and $m=1$ modes in the central
part of galaxies, with some hints of their role in the redistribution of angular momentum,
and more specifically in the triggering of gas infall towards the centre. 
The main points to be remembered from this paper are then:
\begin{itemize}
\item Density waves are present in the central part of galaxies, and
do have an important role in the evolution of the central structures.
\item Gas is a critical component for these modes to persist, as they tend
to dynamically heat the system, with timescales obviously much shorter than
for similar modes in the outer parts of galaxies.
\item $m=1$ may be common, at scales of a few hundreds of pc,
and superimposed on central spirals and/or bars, but also at scales
of a few pc where the potential is dominated by the central mass 
concentration.
\end{itemize}

\end{document}